\documentclass[aps,prl,twocolumn,longbibliography,floatfix]{revtex4-2}
\usepackage{latexsym}
\usepackage{amssymb,amsmath,amsthm, verbatim}
\usepackage{graphicx}
\usepackage{xcolor}
\usepackage{qcircuit}
\usepackage{bm}	
\usepackage{textcomp}
\usepackage{qcircuit}

\newcommand{\ket}[1]{| #1\rangle}
\newcommand{\bra}[1]{\langle #1 |}

\newtheoremstyle{def}{10pt}{20pt}{}{}{\rmfamily\mdseries\scshape}{.}{.5em}{}
\theoremstyle{def}

\begin{document}
\title{Quantum Neuron with Separable-State Encoding}
\author{
London A. Cavaletto$^{1}$,~Luca Candelori$^2$,~Alex Matos-Abiague$^{3}$
}

\affiliation{
$^{1}$\textit{Department of Computer Science, Wayne State University, Detroit, MI 48201, USA}\\
$^{2}$\textit{Department of Mathematics, Wayne State University, Detroit, MI 48201, USA}\\
$^{3}$\textit{Department of Physics \& Astronomy, Wayne State University, Detroit, MI 48201, USA}
}

\date{\today}
\begin{abstract}
The use of advanced quantum neuron models for pattern recognition applications requires fault tolerance. Therefore, it is not yet possible to test such models on a large scale in currently available quantum processors. As an alternative, we propose a quantum perceptron (QP) model that uses a reduced number of multi-qubit gates and is therefore less susceptible to quantum errors in current actual quantum computers with limited tolerance. The proposed quantum algorithm is superior to its classical counterpart, although since it does not take full advantage of quantum entanglement, it provides a lower encoding power than other quantum algorithms using multiple qubit entanglement. However, the use of separable-sate encoding allows for testing the algorithm and different training schemes at a large scale in currently available non-fault tolerant quantum computers. We demonstrate the performance of the proposed model by implementing a few qubits version of the QP in a simulated quantum computer. The proposed QP uses an N-ary encoding of the binary input data characterizing the patterns. We develop a hybrid (quantum-classical) training procedure for simulating the learning process of the QP and test their efficiency. 
\end{abstract}
\maketitle

\section*{Introduction}
Quantum artificial intelligence (QAI) is a newly emerging field, combining concepts from quantum computing and artificial intelligence. Quantum computers use quantum bits (qubits). Unlike conventional bits, which can be in either state 0 or 1 at a time, qubits can be in superposition forming any linear combination of states 0 and 1 simultaneously \cite{QC2000}. The quantum parallelism inherent to qubits can provide an impressive increase in the computational power of a quantum computer compared to conventional computers. In fact, quantum computers are predicted to be able to solve certain classes of problems that are impossible to solve with any, even the most powerful, conventional computers \cite{QC2000}. Recent advances in building quantum computers have made possible to implement quantum algorithms and evaluate their performance \cite{IBMQ}. 

Structurally resembling synaptic neural connections found in biological systems, artificial neural networks are powerful computational models composed by interconnected nodes. These nodes, or artificial neurons, may follow schema from various models, such as the perceptron model \cite{AI2010}. The goal of this research is to explore how to use the potential advantages of quantum computing in the design of quantum neurons as the fundamental elements of quantum artificial neural networks (QANNs) for QAI applications. Accurate and detailed modeling of many biological, chemical, and physical systems in conventional computers is limited by the high complexity of these systems. QAI offers new hope to tackle these problems and could result in multiple technological applications. In this work we focus on the quantum realization of a neuron model known as the perceptron. A quantum perceptron model has recently been realized and executed in IBM quantum computers for the case of a few qubits data set \cite{Tacchino2019:npjQI}. However, this and other previously proposed quantum algorithms \cite{Andrecut2002:IJMP,Schuld2015:PLA,Neukart2014:PE} use a large amount of multi-controlled gates which complexity increases with the number of qubits. In current quantum architectures multi-controlled gates are very sensible to errors. Therefore, the fidelity of previously proposed algorithms rapidly deteriorates when increasing the number of qubits in currently available quantum processors. In contrast, our algorithm uses a single multi-controlled gate, independently of the number of qubits, promising a higher fidelity and a reduced need for error corrections. The price to pay for these advantages is a decrease in the encoding power with respect to other algorithms that fully exploit quantum entanglement.

The following sections describe the realization of a quantum perceptron. The first section explains how one can use qubits to encode a pattern from a two-color pixel image. Following this section is a discussion on the use of a quaternary scheme for multi-qubit pattern encoding before providing details on how to recognize a pattern using single-qubits. The latter section is further extended to the multi-qubit case using the quaternary scheme. Having defined pattern encoding and recognition for multiple qubits, the proposed perceptron algorithm is defined for a quantum machine before concluding with a discussion on results.

\begin{figure*}
	\centering
	\includegraphics*[width=11cm]{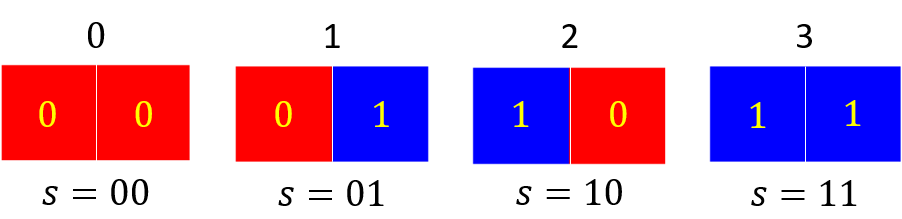}
	\caption{Possible patterns with a two-pixel red and blue image. Red and blue pixels have been encoded into the bit states `0' and `1'. Each pattern is represented by a binary string $s$.}\label{fig1}
\end{figure*}

\section*{N-ary single-qubit pattern encoding}

Suppose there is a two-dimensional image composed of only red and blue pixels, where each pixel is associated a '0' for red or '1' for blue. On a classical computer, each image can be represented by a binary string $S$, where $S = s_0s_1...s_n$. If the image contains two pixels, then the image can take on one of four possible patterns (see Fig.~1). In the general case for an arbitrary $N$-pixel image, there exists $2^N$ distinguishable patterns. 
Unlike classical bits, qubits can be initialized to a superposition of state $\ket{0}$ and $\ket{1}$ \cite{QC2000}. This is given by,
\begin{equation}\label{psi-0}
\ket{\psi}=a\ket{0}+b\ket{1},
\end{equation}
where $a$ and $b$ can be any complex number such that $|a|^2+|b|^2=1$. Noting that $a$ and $b$ can be parametrized further, let these be real numbers and set them as $a=\cos{\theta}$ and $b=\sin{\theta}$. Now Eq.~(\ref{psi-0}) can be rewritten as,

\begin{equation}\label{psi-1}
\ket{\psi}=\cos\theta\ket{0}+\sin\theta\ket{1}.
\end{equation}

Consider the case where one wishes to encode one of the images shown in Fig.~1 using a qubit. By associating each pattern to a unique $\theta\in[0,\pi)$, one can encode the image by using one of four single-qubit states corresponding to $\theta=0,\pi/4,\pi/2,3\pi/4$ and retrieve the corresponding equivalences represented in Table~1.
\begin{table}
\begin{ruledtabular}
\begin{tabular}{cccccc}
		& {\bf Label}& {\bf Angle}  & {\bf Value}  & {\bf State}  &  \\ 
	\hline 
		 & $0$ & $\theta_0$ & $0$ & $\ket{0}$ &\\
	\hline 
		 & $1$ & $\theta_1$ & $\frac{\pi}{4}$ & $\frac{\ket{0}+\ket{1}}{\sqrt{2}}$ &\\
	\hline 
		 & $2$ & $\theta_2$ & $\frac{\pi}{2}$ & $\ket{1}$ &\\
	\hline 
		 & $3$ & $\theta_3$ & $\frac{3\pi}{4}$ & $\frac{\ket{0}-\ket{1}}{\sqrt{2}}$ &\\
\end{tabular} 
\caption{The above table illustrates the quartenary labels and their corresponding angles to generate the single-qubit states.}
\end{ruledtabular}
\end{table}
In theory, there are infinite number of values of $\theta$ between $0$ and $\pi$ and one can encode an infinite number of patterns with a single qubit. However, the number of qubits that may be implemented in practice is limited by the fidelity of the encoding and the error threshold of the quantum processor. For an image of arbitrary size $N$, there are $m=2^N$ possible patterns for a single qubit,
\begin{align*}
\ket{\psi_j}=&\cos\theta_j\ket{0}+\sin\theta_j\ket{1}
\end{align*}
where each qubit is associated an angle,
\begin{align*}
\theta_j=&j\frac{\pi}{m} \;\;\;\;\;\;\;\;\; (j=0,1,...m-1) 
\end{align*}
to the $j$th pattern.

To build the state $\ket{\psi_j}$ encoding the $j$th input pattern, one needs to apply a unitary operation $U_j$ such that
\[ \ket{\psi_j}=U_j\ket{0}. \]
where the operation $U_j$ is defined as follows,
\begin{equation}\label{uj}
U_j = U(\theta_j) = \left(\begin{array}{cc}
\cos(\theta) & -\sin(\theta) \\ 
\sin(\theta) & \cos(\theta)
\end{array} \right).
\end{equation}

\begin{figure*}[t]
	\centering
	\includegraphics*[width=11cm]{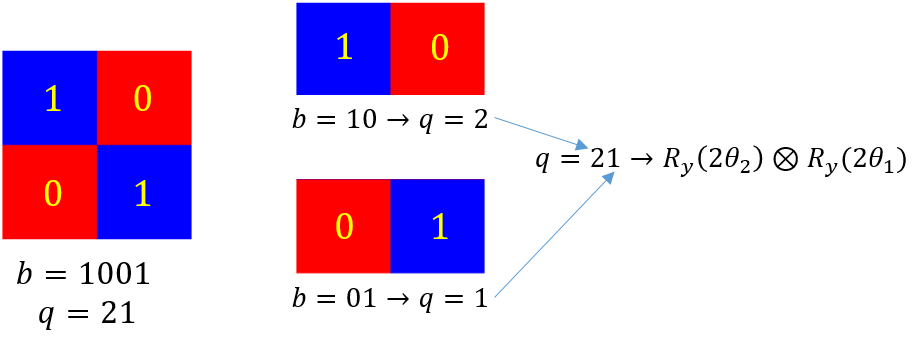}
	\caption{A four-pixel pattern represented by the binary string $b=1001$ can be decomposed into two two-pixel sub-patterns characterized by the binary strings $b=10$ and $b=01$, corresponding to a quaternary string $q=21$. The quaternary string can then be encoded in the two-qubit state $U(2\theta_2)\otimes U(2\theta_1)\ket{00}$.
	}\label{fig2}
\end{figure*}

\section*{Quarternary multi-qubit pattern encoding}

Following the arguments from the previous section, this can extend to the multi-qubit pattern encoding case. Given $M$ bits it is possible to encode $2^M$ patterns on a classical machine. On a quantum machine with a register of the same size, it is possible to encode $N^M$ patterns using an N-ary encoding scheme. Suppose the chosen scheme is for quaternary encoding, then one may encode up to $2^{2N}$ patterns.

Given a two-color image or binary string, one can assign a label `0', `1', `2', and `3' to each pair of consecutive blocks in the image from left to right, up and down as shown in Fig.~1. In case the last pair contains an odd number of bits, a 0 is appended before converting to quaternary. For example, consider a binary string corresponding to a four-pixel image is '1001'. For converting to quaternary, the input is first split into consecutive blocks of pairs '10' and '01' then these are replaced with the corresponding quartenary values '2' and '1' as illustrated in Fig.~2. 

The idea of the quaternary encoding is to set each qubit to encode any of the four elemental patterns in Table~1,
\begin{equation}\label{q-enc}
j\;\;\;\rightarrow\;\;\;U(2\theta_0)\ket{0}\;\;\;,\;\;\;\theta_j=j\frac{\pi}{4}\;\;\;(j=0,1,2,3).
\end{equation}

In the quaternary encoding, the operator $U_j$ can be build by applying the operations given in Eq.~(\ref{q-enc}) for each elemental quaternary state that appears in the quaternary string. Following the previous example, the pattern characterized by the quaternary string q=`21' can be encoded into a quantum register of the same size using the rotations $U(2\theta_2)$ and $U(2\theta_1)$ on the respective qubits. In such a case, we have $U_{21}=U(2\theta_2)\otimes U(2\theta_1)$.

In general, the quaternary encoding of the input pattern into a two-qubit state is constructed as,
\[ \ket{j_0j_1} =U_{j_0j_1}\ket{0}=U(2\theta_{j_0})\otimes U(2\theta_{j_1})\ket{00}, \]
where $j_0$ and $j_1$ are the elements of the quaternary number labeling the input pattern. This scheme can be easily extended to systems with any number of qubits. For an n-qubit register,

\begin{equation}\label{ujn}
\begin{aligned}
\ket{j_0j_1...j_{n-1}} ={}& U_{j_0j_1...j_{n-1}}\ket{0}_n \\
={}&\bigotimes^{n-1}_{i=0}{U(2\theta_{j_i})} \ket{0}_n, 
\end{aligned}
\end{equation}
where $j_0j_1...j_{n-1}$ is the quaternary specifying the input pattern.

\section*{N-ary single-qubit pattern recognition}

Suppose you want to compare an input pattern encoded in the state $\ket{\psi_j}$ against a target pattern encoded in the state $\ket{\psi_t}$. Due to the orthonormality of qubit states, $\bra{\psi_t}\psi_j\rangle=1$ only if $\ket{\psi_j}=\ket{\psi_t}$. Therefore, by measuring the quantity
\begin{equation}\label{prob-0}
P_{tj}=|\bra{\psi_t}\psi_j\rangle|^2
\end{equation}
one can determine whether the input and target patterns are the equal. Using Eqs.~(\ref{psi-0}) and (\ref{prob-0}) yields,
\begin{equation}
\begin{aligned}
P_{tj}={}&|\cos\theta_t\cos\theta_j+\sin\theta_t\sin\theta_j|^2\\
={}&\cos^2(\theta_t-\theta_j).
\end{aligned}
\end{equation}
From this relation it is clear that if the input and target are equal (i.e. $\theta_j=\theta_t$) then $P_{tj}\rightarrow 1$.

\begin{figure}[t]
	$$
	\Qcircuit @C=.6em @R=1em { 
		\lstick{\ket{0}} & \qw & \multigate{3}{U_i} & \multigate{3}{U_w} & \multigate{3}{X^{\otimes N}} & \ctrl{4} & \qw & \qw \\
		\lstick{\ket{0}} & \qw & \ghost{U_i} & \ghost{U_w} & \ghost{X^{\otimes N}} & \ctrl{3} & \qw & \qw \\
		\lstick{\vdots} &  & \nghost{U_i} & \nghost{U_w} & \nghost{X^{\otimes N}} &  & \vdots \\
		\lstick{\ket{0}} & \qw & \ghost{U_i} & \ghost{U_w} & \ghost{X^{\otimes N}} & \ctrl{1} & \qw & \qw \\ 
		\lstick{a = \ket{0}} & \qw & \qw & \qw & \qw & \targ & \meter & \qw \\
		& \cw & \cw & \cw & \cw & \cw & \cw \cwx[-1] & \cw
	}
	$$
	\caption{General circuit diagram for proposed perceptron where $U_i$ sets the register to desired input vector, $U_w$ implements the current weight vector relative to the circuit, parallel $X$-gates } \label{cir-diag}
\end{figure}
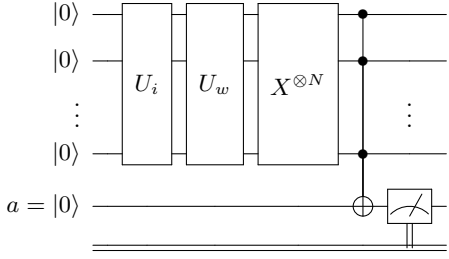

Let $U_j$ and $U_t$ be the unitaries transforming the state $\ket{0}$ into the input and target states, respectively. Then
\[\ket{\psi_j}=U_j\ket{0}\;\;\;,\;\;\; \ket{\psi_t}=U_t\ket{0} \]
and Eq.~(\ref{prob-0}) yields,
\[ P_{tj}=\bra{\psi_t}\psi_j\rangle = \bra{0}U_t^\dagger U_j\ket{0}. \]

Using Eq.~(\ref{uj}) and taking into account that,
\begin{equation}\label{udag}
U_t^\dagger=U(-2\theta_t)
\end{equation}
one obtains,
\begin{equation}\label{uu-0}
\begin{aligned}
U_t^\dagger U_j\ket{0}={}&U(-2\theta_t)U(2\theta_j)\ket{0} \\
={}&\cos(\theta_t-\theta_j)\ket{0}+\sin(\theta_t-\theta_j)\ket{1}.
\end{aligned}
\end{equation}
Therefore, if we measure the qubit state after applying $U_t^\dagger U_j$ to $\ket{0}$ we will find the qubit is still in the $\ket{0}$ state with probability,
\[P_{tj}=\cos^2(\theta_t-\theta_j).\]
If the probability of measuring the $\ket{0}$ is 1 we can conclude that the input and target patterns are the same, otherwise the input differs from the target.

\section*{Quaternary pattern recognition with multiple qubits}

For the pattern recognition, the procedure is similar to the case of N-ary encoding with a single qubit. We need to encode a predefined target state as,

\begin{figure*}[t]
	\centering
	\includegraphics*[width=14cm]{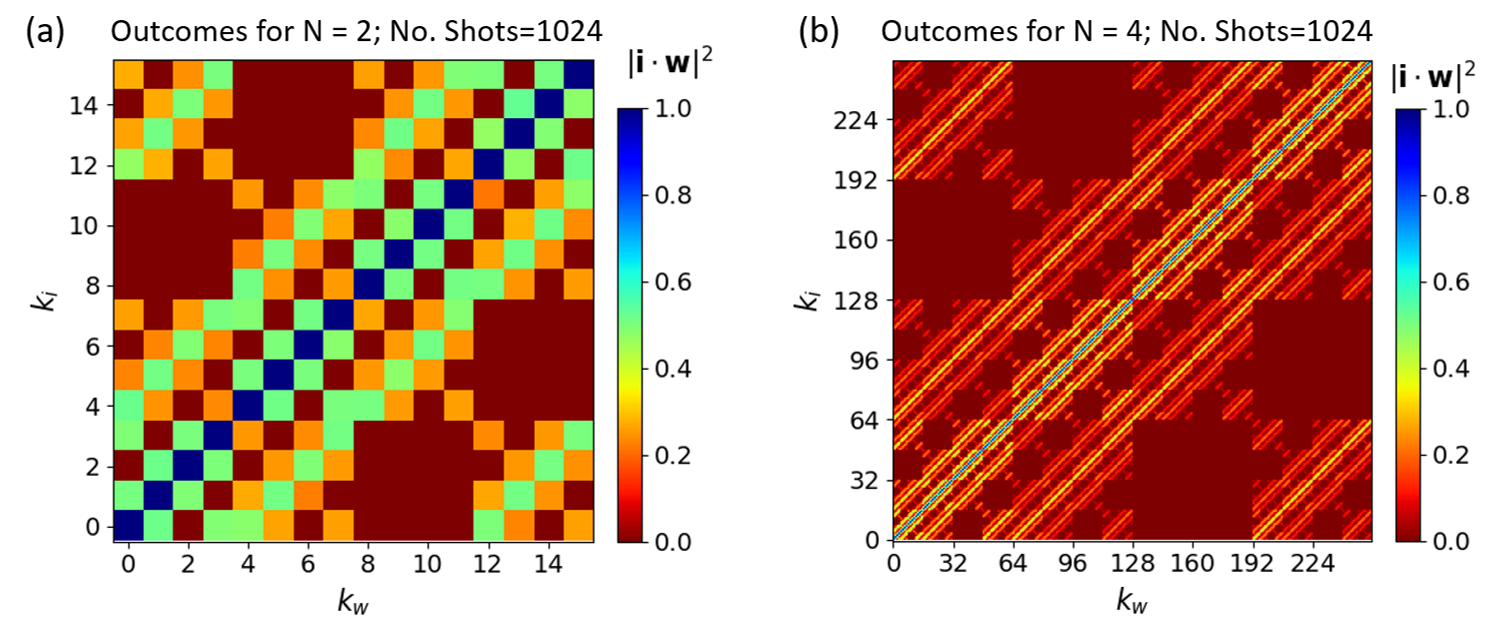}
	\caption{Heatmaps for the scalar product of pairs $(k_i, k_w)$ as a function of all possible inputs and weights, with $k_i$ and $k_w$ representing the decimal representation of the binary strings characterizing the inputs and weights, respectively. The measurements are obtained using IBM's QASM simulator using the quaternary encoding scheme. The plot shown in (a) is given for the N=2 case, while the plot in (b) is given for N=4.}\label{fig:outcomes}
\end{figure*}

\begin{align*}
\ket{t_0t_1...t_{n-1}} =U_{t_0t_1...t_{n-1}}\ket{0}_n
=\bigotimes^{n-1}_{i=0}{U(2\theta_{t_i})\ket{0}_n}.
\end{align*}
where $t_0t_1...t_{n-1}$ is the quaternary labeling the target pattern. The generalization of Eq.~(\ref{uu-0}) to the quaternary case reads,

\begin{equation}\label{uu-1}
\begin{aligned}
\bigotimes^{n-1}_{i=0}{U^\dagger(2\theta_{t_i})}\bigotimes^{n-1}_{i=0}{U(2\theta_{j_i})}\ket{0}_n=\\
=\bra{t_0t_1...t_{n-1}}j_0j_1...j_{n-1}\rangle \ket{0}_n+...
\end{aligned}
\end{equation}
After these operations, the qubits will be found in the state $\ket{00...0}$ with probability,
\[ P_{tj}=|\bra{t_0t_1...t_{n-1}}j_0j_1...j_{n-1}\rangle|^2 ,\]
which equals one only if the input and target patterns are equal.

In order to measure $P_{tj}$ one needs to flip all the qubits, so that one can use the first state on the right-hand side of Eq.~(\ref{uu-1}) as control qubits and add an ancillary qubit to be used as the target. This leads to the following transformations,
\begin{widetext}
\begin{equation}\label{cx-x-uu}
c_{n}X(j_0j_1..j_{n-1},a)(X^{\otimes^N}\otimes\mathbb{I})\bigotimes^{n-1}_{i=0}{U^\dagger(2\theta_{t_i})}\bigotimes^{n-1}_{i=0}{U(2\theta_{j_i})}\ket{0}_n\ket{0}_a
=\bra{t_0t_1...t_{n-1}}j_0j_1...j_{n-1}\rangle \ket{1}_n\ket{1}_a+...
\end{equation}
\end{widetext}
Here the subindex $a$ labels the ancillary qubit, and $c_{n}X(j_0j_1..j_{n-1},a)$ is a multi-controlled NOT gate taking all qubits but the ancilla as the control and the ancilla as the target. The probability of measuring the ancillary qubit in state $\ket{1}_a$ is then equal to $P_{tj}$. For an illustration of the complete set of $P_{tj}$ using the quartenary scheme for N=2 and N=4 qubits, see Fig.~\ref{fig:outcomes}.

\section*{Implementation on a Quantum Processor}
Given the previous sections, it is now possible to construct a circuit for a perceptron on a quantum processor. The quantum register is initialized to a desired size $N+1$, where the extra qubit is reserved for the ancilla bit. $U_i$ is applied to the first $N$ qubtis, as in Eq.~(\ref{ujn}), to construct the input vector. The matrix $U_w$ corresponding to the $w$-vector is constructed as in Eq.~(\ref{udag}) before being applied to the register to obtain the scalar product which is stored in the first index. To retrive the product, a multicontrolled-NOT gate must be applied onto the register to cast the value onto the ancilla bit before measurement (see Fig.~3). Remarkably, this is the only multicontrolled gate required for our quantum algorithm, all the other operations used in the algorithm are single-qubit gates.

Running the circuit $2^n$ times, one can use the ratio of the ancilla measurements to determine the estimated value $\widehat{O}(i,w)$. This value is then compared against the expected value $O(i,w_t)$. If $\widehat{O}(i,w)=O(i,w_t)$, then the perceptron correctly classified the input using the current weight $w$ and there is no need to update $w$. If $\widehat{O}(i,w)\not=O(i,w_t)$, then a mismatch occured and $w$ must be updated using a training procedure. In the latter, three cases are handled by the training procedure: \\

{\bf\emph{Case 1}}. If $O(i,w)=0$ was mesaured when $\widehat{O}(i,w)>0$ or $O(i,w)>0$ was measured when $\widehat{O}(i,w)=0$, then there exists at least one pair of input and weight vector elements $(i_j,w_j)$ that should be orthogonal or that shouldn't be orthogonal. Hence, one can select one random index from $w$ to increased to the next value. \\

{\bf\emph{Case 2}}. If $O(i,w)>\widehat{O}(i,w)$, then the vectors are scanned for indexes containing equal elements, one of these is chosen randomly from $w$ and it increased to the next quaternary value.\\

{\bf\emph{Case 3}}. If $O(i,w)<\widehat{O}(i,w)$, then the vectors are scanned for indexes containing different elements, one of these is chosen randomly from $w$ and it is set the same to the corresponding element in $i$.\\

The training procedure repeats until a cycle is completed without any misclassification errors.

\begin{figure*}
	\centering
	\includegraphics*[width=16cm]{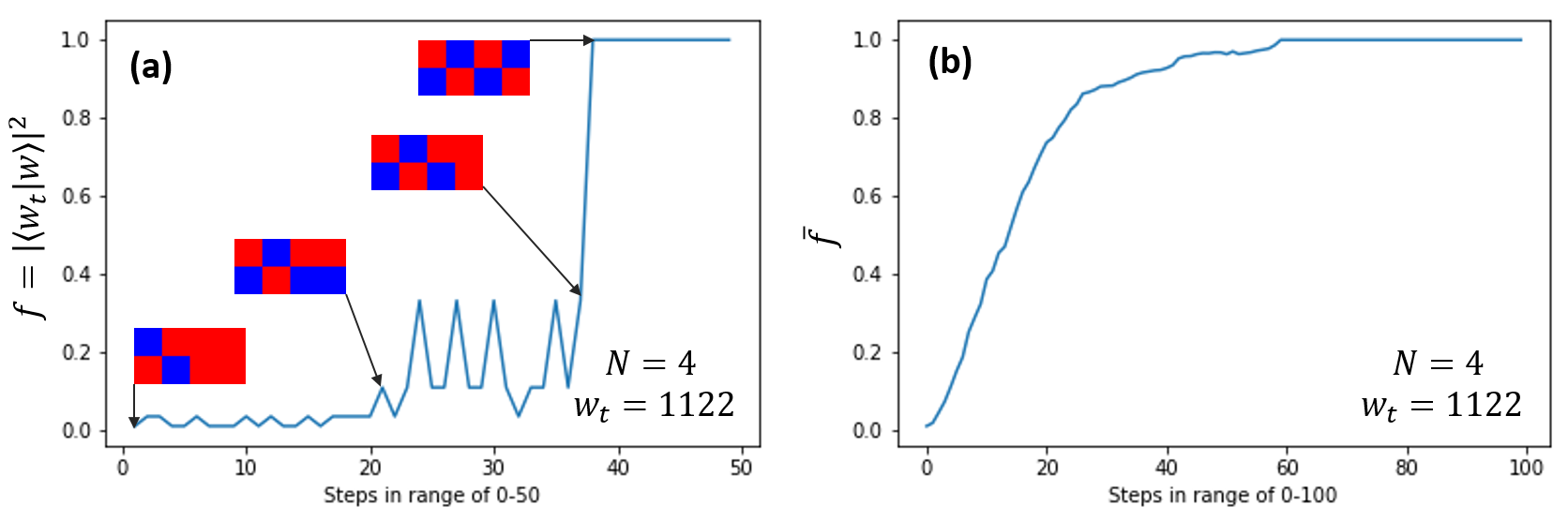}
	\caption{Fidelity as a function of the training steps, for the proposed perceptron using quaternary encoding. The perceptron was run on IBM's simulated quantum computer using 1024 shots. In (a) the training procedure converges after 38 steps using a register with four qubits for the quaternary target $w_t=1122$ (see pattern in the inset). Once the fidelity converges to 1, the quantum perceptron has learned to recognize the target pattern. (b) Fidelity averaged over 200 samples with 1024 shots each.}\label{res}
\end{figure*}

\section*{Results}

In order to evaluate the performance of our quantum algorithm we investigated the capacity of the quantum perceptron to recognize a given, predetermined pattern. This can be characterized by computing the probabilities, $P_{iw}=|\mathbf{i}\cdot\mathbf{w}|^2$ of all the possible outcomes as a function of input and weight patterns. The results are shown in Fig.~4, where $k_i$ and $k_w$ denote the decimal representation of the binary string characterizing the state of each pixel in the input and weight patterns, respectively. For each possible input pattern there is always an identical weight pattern that occurs with high probability (see values along the plots' diagonal), while all the other combinations occur with probability equal or smaller than $0.5$. This large contrast between the values of the probabilities provides a good margin of reliability even in the presence of moderate quantum errors. Consequently, the quantum perceptron can recongnize a predertimed target pattern with very high efficiency. Figures 4(a) and 4(b) correspond to the cases of $N=2$ and $N=4$ qubits, respectively. The probabilities were obtained after 1024 repetitions (shots) of the experiment. Notably the probability along the diagonal remains close to 1 even when the number of qubits increases. Although we do not show the results here, we have checked that this trend persist for $N>4$ too.

The learning ability of the quantum perceptron to recognize a predetermined target pattern after a training is characterized by the improvement of the fidelity,
\begin{equation}\label{fid}
f=|\bra{w_t}w\rangle|^2,
\end{equation}
as the training evolves. The fidelity constitutes a measure of the ability of the perceptron to adjust its weight to the target weight value. When $f=1$, the perceptron can perfectly recognize the given pattern. The dependence of the fidelity on the number of steps during a learning session is shown in Fig.~5 for the case $N=4$. The fidelity at each training step was averaged over 1024 shots. The initial weight was chosen randomly and is shown in Fig.5(a) as the pattern at step 0. As the learning process advances the weight patterns are gradually updated. Some weight patterns at different learning steps are also shown for illustration. By the end of the learning training [step 38 in Fig.5(a)] the perceptron has learned to properly recognize the checkerboard pattern, characterized by the quaternary string ``1122" and chosen as the target. Since the number of learning steps the perceptron needs to complete the training can vary from session to session, we have also investigated the average fidelity over 200 uncorrelated training sessions. The results are shown in Fig.5(b), where one can see that, on average, the perceptron needs less than 60 steps to perfectly recognize a pattern.

Although the results in Fig.~5(a) are satisfactory, the oscillatory behavior of the fidelity indicates that there is still room for improvement. In fact, preliminary results (not shown here) indicate that using equiangular basis for the N-ary encoding may not only improve the efficiency of the training procedure, but may also allow for extending classification beyond the binary case. A hybrid approximation, where the encoding power is largely increased by combining multiple-qubit perceptrons with N-ary encoding and entangled states used on demand will be discussed elsewhere \cite{Cavaletto2022:U}.

\section*{Conclusion}
We developed a new quantum algorithm capable of simulating a perceptron. The algorithm allows for an exponential increase in the number of patterns that can be encoded per qubit, compared to classical perceptron models. Unlike previously proposed schemes, our perceptron implementation uses mainly separable-state and single-qubit operations, potentially leading to a considerable reduction of quantum error effects. The proposed quantum algorithm uses N-ary encoding to increase the density of encoded pixels per qubit. In particular, the performance of a quantum perceptron with quaternary encoding was investigated by using the facilities of the IBM Quantum Experience platform \cite{IBMQ}. The functionality of the proposed quantum perceptron for pattern recognition was tested by creating and implementing a training procedure through which the perceptron learned to correctly identify a target pattern. The encoding power could be largely increased by combining the N-ary encoding with the use of entangled states according to demand and the specific quantum processor capabilities.

\section*{Acknowledgments} 
L.A.C acknowledges support by UROP Wayne State University.

\section*{References}

\bibliography{BibToQC}

\end{document}